\documentclass[conference]{IEEEtran}

\usepackage{hyperref}
\usepackage{url}
\usepackage{booktabs}
\usepackage{tabularx}
\usepackage{color}
\usepackage{graphicx}

\usepackage{multirow}
\usepackage{amsmath}
\usepackage{amssymb}
\usepackage{rotating}
\usepackage[linesnumbered, ruled]{algorithm2e}
\usepackage{balance}

\newcommand{\etal}{\textit{et al. }}

\ifCLASSINFOpdf
\else
\fi
\begin{document}

%
\title{Assessing and Improving the \\Mutation Testing Practice of PIT}



%
\author{\IEEEauthorblockN{Thomas Laurent\IEEEauthorrefmark{1}\IEEEauthorrefmark{2},
Anthony Ventresque\IEEEauthorrefmark{1},
Mike Papadakis\IEEEauthorrefmark{3}, 
Christopher Henard\IEEEauthorrefmark{3}, and
Yves Le Traon\IEEEauthorrefmark{3}}
\IEEEauthorblockA{\IEEEauthorrefmark{1}Lero@UCD, School of Computer Science, University College Dublin, Ireland}
\IEEEauthorblockA{\IEEEauthorrefmark{2}Ecole Centrale de Nantes, France}
\IEEEauthorblockA{\IEEEauthorrefmark{3}Interdisciplinary Centre for Security, Reliability and Trust, University of Luxembourg, Luxembourg}
\IEEEauthorrefmark{2}thomas.laurent@eleves.ec-nantes.fr, \IEEEauthorrefmark{1}anthony.ventresque@ucd.ie, \IEEEauthorrefmark{3}\{firstname.lastname@uni.lu\}}


\maketitle

\begin{abstract}
Mutation testing is used extensively to support the experimentation of software engineering studies. Its application to real-world projects is possible thanks to modern tools that automate the whole mutation analysis process. However, popular mutation testing tools use a restrictive set of mutants which do not conform to the community standards as supported by the mutation testing literature. This can be problematic since the effectiveness of mutation depends on its mutants. We therefore examine how effective are the mutants of a popular mutation testing tool, named PIT, compared to comprehensive ones, as drawn from the literature and personal experience. We show that comprehensive mutants are harder to kill and encode faults not captured by the mutants of PIT for a range of 11\% to 62\% of the Java classes of the considered projects.


\end{abstract}


%
\IEEEpeerreviewmaketitle

\section{Introduction}

Software testing constitutes the current practice for checking programs. 
In such a scenario, sets of test cases are selected and used to examine the behavior of the programs under investigation. 
To quantify the ``quality'' of the test cases, researchers and practitioners use the so-called adequacy metrics or testing criteria \cite{ZhuHM97}. These metrics measure the quality achieved by the employed test sets. 

Mutation analysis is an established test criterion \cite{ZhuHM97, 0020331} that promises to thoroughly examine the programs under investigation. It operates by evaluating the ability of the candidate test cases to distinguish between the program under test and a set of altered program versions, called mutants. Mutants represent program defects and are used to measure the ability of the test cases to reveal them. The power of the technique is based on the ability of the mutants to represent real faults \cite{AndrewsBL05, AndrewsBLN06} and to lead testers in writing test cases that cover almost all the other test criteria \cite{Offutt11, FranklWH97, OffuttPTZ96}.  

The downfall of mutation is its application cost. This is related to the number of possible mutants which can be prohibitively high \cite{5487526, OffuttLRUZ96}. Each mutant forms a different program version that needs to be executed with the candidate test cases. Therefore, a large number of test executions is required in order to compute the adequacy measurement. 

Because of the large number of mutants, practitioners believed that mutation does not scale to real-world programs. However, modern tools proved this belief wrong and as a result mutation ``entered the mainstream'' of practice \cite{Offutt11}. To this end, several tools have been developed, linked with build systems and development tools. Modern tools are also robust and they can easily be used by developers \cite{DelahayeB15}. As a result, they are used extensively in software engineering studies. 

Unfortunately, popular tools like PIT \cite{pit} employ a restrictive set of mutants that does not fully conform to the recommendations made by the mutation testing literature. This fact indicates potential issues with the effectiveness of the tools given that mutation is sensitive to its mutants \cite{NaminK11}. Since these tools are extensively used in software engineering studies, it is mandatory to validate the extent to which their adopted mutant set is representative of the community standards as supported by the mutation testing literature.

This paper presents a thorough study investigating the above-mentioned issue using PIT \cite{pit}. We use PIT since it was found to be the most robust available mutation testing tool \cite{DelahayeB15} and it has been used extensively for research purposes in the recent years, e.g., \cite{ZhangM15, DenaroMPV15, InozemtsevaH14, ShiGGZM14}. Our study indicates significant limitations of the popular mutants and thus, motivate the need for a more comprehensive one.

\noindent In summary, the contributions of this paper are:

\begin{itemize}

\item We describe and implement a comprehensive mutant set for Java that  reflects the beliefs of the mutation testing community as it has been recorded in the literature \cite{OffuttLRUZ96, AndrewsBLN06} and discussed during the Mutation 2014 and 2015 workshops, e.g., \cite{MutKeynote2015}. 
\item We provide empirical evidence that the comprehensive mutant set is superior to the one often used by mutation testing tools. This set is statistically significant superior from 11\% to 62\% of the studied program classes. 
\item To support future research, we will submit our code to the PIT repository to became available. Our new version of PIT that supports the comprehensive mutants is also available on request\footnote{For inquires please contact Anthony Ventresque, anthony.ventresque@ucd.ie}. 
\end{itemize}







\section{Terminology \& Background}\label{bg}
This section introduces the terminology and the concepts that are used throughout the paper. First,  \ref{sec:mut} presents the mutation testing process. Then, Section \ref{sec:red} describes the selection of mutants. Finally, the notion of disjoint mutants is introduced in Section \ref{sec:disjoint}.


\subsection{Mutation Testing}\label{sec:mut} 

Mutation analysis operates by injecting defects in the software under investigation. Thus, given a program, several variants of this program are produced, each variant containing a defect. These are called \textit{mutants} and they are made by altering (mutating) the code, either source code or executable binary code, of the program under test. The creation of mutants is based on syntactic rules, called \textit{mutant operators}, that transform the syntax of the program. For example, an arithmetic mutant operator changes an instance of an arithmetic language operator such as the '$+$' to another one, such as '$-$'. 


Mutants are produced by syntactic changes introduced by mutant operators. These are the instances that are produced by applying an operator on every point of the code under investigation that matches their respective rule. Thus, every mutant has a single and specific syntactic difference from the original program. For reasons that we will discuss in the related work (see Section \ref{related}), mutation testing uses mutants produced by simple syntactic changes. 

Mutants are used to measure how good the employed test cases are in checking the software under assessment. This is done by observing the runtime behavior of the original, non-mutated, and the mutated programs. When comparing the program outputs of the original with the mutated programs, and found differences, we exhibit behavior discrepancies \cite{PapadakisT15}. Such differences are attributed to the ability of the used test to project the syntactic program changes to its behavior, i.e., to show a semantic difference. When mutants exhibit such differences in their behavior, they are called ``killed''. Those that do not exhibit such differences are called ``live''. Mutants might not exhibit any difference in their behavior either because the employed test cases were not capable of revealing them or because they are functionally equivalent with the original program. Mutants belonging to the latter case are called equivalent \cite{PapadakisT15}. 
 
Mutation testing refers to the process of using mutation analysis as a means of quantifying the level of thoroughness of the test process. Thus, it measures the number of mutants that are killed and calculates the ratio of those over the total number of mutants. This ratio represents the adequacy metric and is called mutation score. Ideally, to have an accurate metric, equivalent mutants must be removed from the calculation of the score. However, this is not possible since judging programs' equivalence is an undecidable problem \cite{BuddA82}.

%
%
%

\subsection{Mutant Selection}\label{sec:red}
%


Selective mutation was shown to be valid in several studies involving programs written in Fortran \cite{OffuttLRUZ96} and in C \cite{VincenziMBD01}. As a result, Java mutation tools were built based on the findings of these studies. To address the scalability issues of the method, tool developers made further reductions. Thus, popular tools like 
 PIT \cite{pit} support a very small and restrictive set of mutants that neither follows any suggestion from previous studies nor practical experience. 

PIT, even in its latest version that supports an extended mutant set, has several shortcomings. One such example is the relational operator for which PIT 
replaces one instance of the operator by only another one, i.e., mutates $<$ only to $<=$, or $<=$ only to $<$, or $>$ only to $>=$, or $>=$ only to $>$. However, this practice is not sufficient. Indeed, previous studies have shown that three mutants are needed to avoid a reduced effectiveness of the method \cite{Tai06, PapadakisBM10}. 


Although practical, mutant reduction should not be at the expense of the method effectiveness. Almost all previous studies were based on the assumption that mutants are equal \cite{JiaH09a}. However, this does not hold in practice and has the potential to bias the conducted research as recent studies show \cite{AmmannDO14, JiaH09a, KintisPM10}. Therefore,  when using mutation for research purposes, it is mandatory to make sure that a representative mutant set is employed. 


\subsection{Disjoint Mutants}\label{sec:disjoint}

In literature, mutation testing is extensively used to support experimentation \cite{AndrewsBL05, 5487526}, i.e., it is used to measure the level of test thoroughness achieved by various testing methods. Mutation score serves as a comparison basis between testing techniques and hence, as a yardstick to judge the winning one. This practice is quite popular, and introduce severe problems that can threaten the validity of the conducted research. 

The problem is that not all mutants are of equal power \cite{JiaH09a}, which means that some are useful and some are not. Indeed, mutants cover the full spectrum of cases, including trivial ones, very easy to kill, duplicated, equivalent ones and also hard to kill ones. Those of the last category are of particular interest since they lead to strong tests \cite{AndrewsBLN06, JiaH09a}. Hard to kill, trivial and easy to kill mutants are defined with respect to the employed test suite \cite{JiaH09a}. Thus, mutants killed by a small percentage of tests that exercise them are hard to kill, while, those killed by a large one are easy to kill. 

When using mutation as a basis for comparing testing methods, a filtering process that sweeps out the duplicated and equivalent mutants is needed \cite{PapadakisJHT15}. However, this process might not be adequate since in most cases many mutants tend to be killed jointly \cite{KintisPM10}. Thus, they do not contribute to the test process despite being considered. This has an inflation effect on the mutation score computation since only a very small fraction of mutants contribute to the test process\footnote{Kintis \etal \cite{KintisPM10} reports that this is 9\% of mutants, for Java programs using the muJava mutation testing tool, Amman \etal \cite{AmmannDO14} report 10\% for the Java mutants of muJava tool and 1\% for the C mutants of the Proteum tool.}. 

This issue was initially raised by Kintis \etal \cite{KintisPM10} who introduced the concept of \textit{disjoint mutants}, i.e., minimum number of mutants that contribute to mutation score. Their use is motivated by the same study which demonstrated that hard to kill mutants also suffer from the inflation problem. Later Amman \etal \cite{AmmannDO14} formalized this concept, name it as ``minimum mutants'', and suggested using it as a way to bypass the mutation score inflation problem.

\begin{algorithm}[!t]

\DontPrintSemicolon
\SetInd{0.4em}{0.4em}
\scriptsize
\KwIn{A set $S$ of mutants}
\KwIn{A set $T$ of test cases}
\KwIn{A matrix $M$ of size $|T|\times |S|$ such as $M_{ij} = 1$ if test$_i$ kills mutant$_j$}
\KwOut{The disjoint mutant set $D$ from S}
$D=\emptyset$\;\tcc*[c]{Remove live mutants}
$S = S \setminus \{m \in S\,|\,\forall i \in 1..|T|, M_{ij} \ne 1 \}$\;\tcc*[c]{Remove duplicate mutants}
$S = S \setminus \{m \in S\,|\,\exists m'\in S\,|\, \forall i \in 1..|T|, M_{ij(m)} = M_{ij(m')} \}$\;
\While{$\left(|S| > 0\right)$}
{
maxSubsumed = 0\;
subsumedMut = null\;
maxMutSubsuming = null\;

\tcc*[c]{Select the most subsuming mutant}
\ForEach{$\left(m \in S\right)$}
{
    $\text{sub}_m = \{m' \in S | \forall i \in 1..|T|, (M_{ij(m)} = 1) \Rightarrow (M_{ij(m')} = 1)\}$\;
    \If{\upshape$\left(|\text{sub}_m| > \text{maxSubsumed}\right)$}
    {
	maxSubsumed $ = |$sub$_m|$\;
	maxMutSubsuming = $m$\;
	subsumedMut = sub$_m$\;
    }
}

\tcc*[c]{Add the most subsuming mutant to $D$}
$D = D \cup \{\text{maxMutSubsuming}\}$\;
\tcc*[c]{Remove the subsumed mutants from the remaining}
$S = S \setminus \text{subsumedMut}$\;

}
return $D$\
\caption{Disjoint Mutants}
\label{algo_disjoint}
\end{algorithm}

In this paper we follow an analysis based on both all and disjoint mutants. 
Disjoint mutants have the advantage to cover the whole spectrum of mutants and suffer less from the mutant inflation effect. Their identification is an NP-complete problem \cite{AmmannDO14} and thus, we use a greedy approximation method. Algorithm \ref{algo_disjoint} details their computation from a set of mutant $S$. First, the live and duplicate mutants are removed from $S$ (lines 2 and 3). Then, the most subsuming mutant is retrieved (lines 8 to 15). It is the mutant which, when killed, implies the highest number of other mutants to be killed as well. This mutant is then added to the disjoint set $D$ (line 16) and the subsumed mutants are removed from $S$ (line 17). This process is repeated until $S$ is empty. Finally, the set of disjoint mutants, $D$, is returned.

%
%
%
%
%
%
%
%

\section{Motivation}\label{prob}
Mutation testing is extensively used by researchers and has an increasing use by practitioners and the open source community \cite{5487526}, mainly due to the existence of automated tools. However, mutation is sensitive to the set of mutants that are used \cite{NaminK11}. Therefore, it is mandatory to equip these tools with a comprehensive set of mutants that can adequately measure test thoroughness.

In this paper, we deal with this issue by investigating the extent to which the mutant sets employed by popular mutation testing tools meet the standards as expressed by the mutation testing literature and community. We call the first set as the ``common'' mutant set and the second one as the  ``comprehensive'' one. 


Our goal is to validate the use of the popular mutant set which was introduced by the developers of popular mutation testing tools. We seek to investigate this issue since modern tools like 
PIT \cite{pit} have been extensively used in the recent years\footnote{For instance, \cite{ZhangM15, DenaroMPV15, InozemtsevaH14, ShiGGZM14} are recent publications that use PIT.}. Thus, a possible issue with their adopted mutants can question the effectiveness of the mutation method and hence the conducted research. We therefore compare the extent to which the popular mutant set conforms to the test requirements possessed by the mutation testing literature. To validate this practice we use large open source projects written in Java with mature test suites.


\section{Experimental Study}\label{xp}

This section first states the Research Questions (RQs) under investigation. Then, the subjects, settings and tools used for the experiments are described. Finally, the last subsections detail the studied mutant operators and the analysis procedure followed to answer the RQs.

\subsection{Definition of the Experiment and Research Questions}

Current research on software engineering has largely focused on using mutation analysis as supported by the existing mutation testing tools. However, a central role in mutation testing is played by the mutants that are used; meaning that the effectiveness of the method is sensitive to the employed mutants \cite{NaminK11}. Therefore, it is important to know whether the commonly used mutants, as supported by these tools, are suitable. In other words we seek to determine the degree to which the commonly used mutants are representative of those suggested by the literature, i.e., the comprehensive mutant set. This leads us to our first research question:
\begin{description}
\item[\textbf{RQ1}] \textbf{(Effectiveness)}. Is there any effectiveness difference between the commonly used mutants and the comprehensive ones? 
\end{description}

 Since we are interested in testing, we seek to identify the mutants that are more effective at measuring the ability of test cases to exercise each point of the program under investigation. 
 Mutation score measurements can differ when different mutant sets are employed. The measurements are affected by the number of: mutants, of equivalent ones, of trivial and hard to kill ones. To deal with this issue, we perform an objective comparison, i.e., we measure the extent to which one method covers the requirements of the other, between the two examined mutant sets. Thus, we seek to measure the ratio of mutants, of the one set, that are found by the tests that are selected based on the other mutant set. Thus, the "weaker" mutants will lead to "weaker" tests and hence kill a smaller fraction of the "stronger" mutants. Here, it should be noted that objective comparisons form a common practice in mutation testing literature, e.g., \cite{AndrewsBLN06, FranklWH97, OffuttLRUZ96}. 
 
  As discussed previously, in Section \ref{sec:disjoint}, there is a potential problem with this practice due to the inflation effect of the trivial mutants. We thus measure the ratios of the disjoint mutants that are killed. Therefore, to answer RQ1 we report results based on two effectiveness measures; the percentage of all mutants killed and the  percentage of the disjoint ones that are killed. 
  
  So far, our investigations focus on whether mutants of one set can capture all the faults introduced by the other set. However, this analysis tells us nothing about the difficulty of exposing mutants. Thus, mutant easiness is another important attribute of mutants \cite{AndrewsBLN06, JiaH09a}. This is due to the fact that hard to kill mutants indicate a relatively small semantic change difference \cite{JiaH09a, OffuttH96} that is often easy to overlook when testing. Thus, through our second research question, we investigate whether the comprehensive mutant set introduces harder to kill mutants than the commonly used one:
      
\begin{description}
\item[\textbf{RQ2}] \textbf{(Easiness)}. What is the difference, in terms of difficulty to expose mutants, between the common and the comprehensive mutants? 
\end{description}
%
%
%
%
%

Mutation testing has a widespread reputation of being computationally demanding. In the past, practitioners believed that it cannot scale to real-world systems mainly due to the large number of mutants. However, tools like PIT and Javalanche proved that this belief was incorect \cite{SchulerZ09}. This ability of the tools can be attributed to the restricted mutant set they employ and to the advanced mutant generation and execution techniques used. Therefore, it is possible that the comprehensive mutant set is too expensive to be used in practice. Hence, we investigate: 

\begin{description}
\item[\textbf{RQ3}] \textbf{(Scalability)}. What is the execution time differences of the comprehensive mutants when compared with the common ones? 
\end{description}

We measure execution time since it forms a direct measure of the application cost of the method. We do not consider other parameters that can influence the application cost, such as the number of equivalent mutants, the test generation cost, since they fall outside the scope of the present paper. Here, we focus on the effectiveness of the method as conducted by recent studies and thus, leaving the issue of its application cost open for future research.  

%


\subsection{Subject Programs}

The experiments are conducted on the 5 Java projects recorded in Table \ref{subjects}. For each of them, the version, lines of code (calculated with the JavaNCSS tool \cite{javancss}), number of classes (for which test suites exist) and number of tests are reported. 

Joda-time is a date and time manipulation library. Jfreechart is a popular library for creating charts and plots. Jaxen is an engine for evaluating XPath expressions. Commons-lang provides a set of utility methods for the commons classes of Java. Finally, commons-collections provides data structures in addition to those existing in the standard Java framework.

%

\begin{table}[!t]
\caption{Subjects used in the experiments. The reported Lines of Code (LoC) and classes are only those corresponding to classes having test cases.}
\label{subjects}
\centering
\begin{tabular}{llrrr}
\toprule
\textbf{Subjects} & \textbf{Version} & \textbf{LoC} & \textbf{Classes} & \textbf{Tests}\tabularnewline 
\midrule 
{joda-time } & 2.8.1 & 18,611 & 210 & 4,129\tabularnewline
{jfreechart} & 1.0.19 &46,986 & 290 & 1,320\tabularnewline
{jaxen } & 1.1.6 &6,790 & 152 & 646\tabularnewline
{commons-lang } & 3.3.4 & 16,286 &199 & 3,373\tabularnewline
{commons-collections } & 4.4.0 & 11,281& 243 & 2,210\tabularnewline
\bottomrule
\end{tabular}
\end{table}

\begin{table*}[!ht]
\setlength{\tabcolsep}{2px} 
\caption{Common and comprehensive mutants.}
\label{tab:ops}
\centering
{\scriptsize{}}%
\begin{tabularx}{\textwidth}{Xl>{\raggedright}m{4.5cm}ll>{\raggedright}m{4.5cm}>{\raggedright}m{2cm}}
\toprule
 & \textbf{\scriptsize{Name}} & \textbf{\scriptsize{Transformation}} & \textbf{\scriptsize{Example}} & \textbf{\scriptsize{Name}} & \textbf{\scriptsize{Transformation}} & \textbf{\scriptsize{Example}}\tabularnewline
\hline 
\multirow{7}{*}{\begin{sideways}
\textbf{\scriptsize{Common op. $\,\,\,\,\,\,\,\,\,\,\,\,\,\,\,\,\,$}}
\end{sideways}} & \textbf{\scriptsize{Cond. Bound. }} & {\scriptsize{Replaces one relational operator instance with another
one (single replacement).}} & {\tiny{$<\,\rightsquigarrow\,\leq$}} & \textbf{\scriptsize{Return Values}} & {\scriptsize{Transforms the return value of a function (single replacement).}} & \texttt{\tiny{return 0$\,\rightsquigarrow\,$return 1}}\tabularnewline
 & \textbf{\scriptsize{Negate Cond.}} & {\scriptsize{Negates one relational operator (single negation).}} & {\tiny{$\ensuremath{=\!=\,\rightsquigarrow\,!\!=}$}} & \textbf{\scriptsize{Void Meth. Call}} & {\scriptsize{Deletes a call to a void method.}} & \texttt{\tiny{void m()$\,\rightsquigarrow\,$}}\tabularnewline
 & \textbf{\scriptsize{Remove Cond.}} & {\scriptsize{Replaces a cond. branch with true or false.}} & \texttt{\tiny{if (...)$\,\rightsquigarrow\,$if (true}}{\tiny{)}} & \textbf{\scriptsize{Meth. Call}} & {\scriptsize{Deletes a call to a non-void method.}} & \texttt{\tiny{int m()$\,\rightsquigarrow\,$}}\tabularnewline
 & \textbf{\scriptsize{Math}} & {\scriptsize{Replaces a numerical op. by another one (single replacement).}} & {\tiny{$+\,\rightsquigarrow\,-$}} &  \textbf{\scriptsize{Constructor Call}} & {\scriptsize{Replaces a call to a constructor by null.}} & \texttt{\tiny{new C()$\,\rightsquigarrow\,$null}}\tabularnewline
 & \textbf{\scriptsize{Increments}} & {\scriptsize{Replace incr. with decr.  and vice versa (single replacement).}} & {\tiny{$\ensuremath{+\!+\,\rightsquigarrow\,-\!-}$}} &\textbf{\scriptsize{Member Variable}} & {\scriptsize{Replaces an assignment to a variable with the Java default
values.}} &  \texttt{\tiny{a = 5$\,\rightsquigarrow\,$a}}{\tiny{}}\tabularnewline
 & \textbf{\scriptsize{Invert Neg.}} & {\scriptsize{Removes the negative from a variable.}} & {\tiny{$-a\,\rightsquigarrow\, a$}} & \textbf{\scriptsize{Switch}} & {\scriptsize{Replaces switch statement labels by the Java default
ones.}} & \tabularnewline
 & \textbf{\scriptsize{Inline Const.}} & {\scriptsize{Replaces a constant by another one or increments it.}} & {\tiny{$1\,\rightsquigarrow\,0,$ $a\,\rightsquigarrow\, a+1$}} & & & \tabularnewline
\hline 
\multirow{4}{*}{\begin{sideways}
\textbf{\scriptsize{Comprehensive op.$\,\,$}}
\end{sideways}} & \textbf{\scriptsize{ABS}} & {\scriptsize{Replaces a variable by its negation.}} & {\tiny{$a\,\rightsquigarrow\,-a$}} & \textbf{\scriptsize{OBBN}} & {\scriptsize{Replaces the operators \& by $|$ and vice versa.}} & {\tiny{$a \& b\,\rightsquigarrow\, a | b$}}\tabularnewline
 & \textbf{\scriptsize{AOD}} & {\scriptsize{Replaces an arithmetic expression by one of the operand.}} & {\tiny{$a+b\,\rightsquigarrow\, a$}} & \textbf{\scriptsize{ROR}} & {\scriptsize{Replaces the relational operators with another one. It
applies every replacement.}} & {\tiny{$<\,\rightsquigarrow\,\geq$, $<\,\rightsquigarrow\,\leq$}}\tabularnewline
 & \textbf{\scriptsize{AOR}} & {\scriptsize{Replaces an artihmetic expression by another one.}} & {\tiny{$a+b\,\rightsquigarrow\, a*b$}} & \textbf{\scriptsize{UOI}} & {\scriptsize{Replaces a variable with a unary operator or removes an
instance of an unary operator.}} & {\tiny{$a\,\rightsquigarrow\, a\!+\!+$}}\tabularnewline
 & \textbf{\scriptsize{CRCR}} & {\scriptsize{Replaces a constant $a$ with its negation, or with  $1$, $0$, $a+1$,
$a-1$.}} & {\tiny{$a\,\rightsquigarrow\, -a,$ $a\,\rightsquigarrow\, a-1$.}} & \textbf{\scriptsize{Commons}}&  \multicolumn{2}{l}{\scriptsize{\textit{All the common operators as described above.}}}\tabularnewline
\bottomrule
\end{tabularx}
\end{table*}

\begin{table*}[!ht]

\caption{Number of mutants, killable mutants and mutation score (MS) for the common and comprehensive mutants.}
\label{operators_table}
\centering


\setlength{\tabcolsep}{3.5px}

{\scriptsize{}}%
{\scriptsize{}}%
\begin{tabular}{lcrrrrrrrrrr}
\toprule
\multicolumn{2}{c}{} & \multicolumn{2}{c}{\textbf{\scriptsize{joda-time}}} & \multicolumn{2}{c}{\textbf{\scriptsize{jfreechart}}} & \multicolumn{2}{c}{\textbf{\scriptsize{jaxen}}} & \multicolumn{2}{c}{\textbf{\scriptsize{commons-lang}}} & \multicolumn{2}{c}{\textbf{\scriptsize{commons-collections}}}\tabularnewline
\multicolumn{2}{c}{\textbf{\scriptsize{Measure}}} & \textbf{\scriptsize{Common op.}} & \textbf{\scriptsize{Compre. op.}} & \textbf{\scriptsize{Common op.}} & \textbf{\scriptsize{Compre. op.}} & \textbf{\scriptsize{Common op.}} & \textbf{\scriptsize{Compre. op.}} & \textbf{\scriptsize{Common op.}} & \textbf{\scriptsize{Compre. op.}} & \textbf{\scriptsize{Common op.}} & \textbf{\scriptsize{Compre. op.}}\tabularnewline
\cmidrule(l{5pt}r{4pt}){1-2} \cmidrule(l{5pt}r{4pt}){3-4} \cmidrule(l{5pt}r{4pt}){5-6} \cmidrule(l{5pt}r{4pt}){7-8} \cmidrule(l{5pt}r{4pt}){9-10} \cmidrule(l{5pt}r{4pt}){11-12}

\multirow{4}{*}{\textbf{\scriptsize{\#Mutants}}} & \textit{\scriptsize{Min.}} & {\scriptsize{1.00}} & {\scriptsize{1.00}} & {\scriptsize{1.00}} & {\scriptsize{1.00}} & {\scriptsize{1.00}} & {\scriptsize{1.00}} & {\scriptsize{1.00}} & {\scriptsize{1.00}} & {\scriptsize{1.00}} & {\scriptsize{1.00}}\tabularnewline
 & \textit{\scriptsize{Med.}} & {\scriptsize{97.00}} & {\scriptsize{224.00}} & {\scriptsize{98.00}} & {\scriptsize{260.50}} & {\scriptsize{24.00}} & {\scriptsize{39.00}} & {\scriptsize{27.00}} & {\scriptsize{57.00}} & {\scriptsize{27.00}} & {\scriptsize{42.00}}\tabularnewline
 & \textit{\scriptsize{Mean}} & {\scriptsize{164.17}} & {\scriptsize{462.06}} & {\scriptsize{219.14}} & {\scriptsize{685.48}} & {\scriptsize{77.48}} & {\scriptsize{188.97}} & {\scriptsize{156.82}} & {\scriptsize{457.05}} & {\scriptsize{62.32}} & {\scriptsize{126.53}}\tabularnewline
 & \textit{\scriptsize{Max.}} & {\scriptsize{973.00}} & {\scriptsize{2,915.00}} & {\scriptsize{3,436.00}} & {\scriptsize{9,742.00}} & {\scriptsize{3,901.00}} & {\scriptsize{14,493.00}} & {\scriptsize{4,545.00}} & {\scriptsize{14,586.00}} & {\scriptsize{1,094.00}} & {\scriptsize{2,349.00}}\tabularnewline[1ex]
\multirow{4}{*}{\textbf{\scriptsize{\#Killable}}} & \textit{\scriptsize{Min.}} & {\scriptsize{0.00}} & {\scriptsize{0.00}} & {\scriptsize{0.00}} & {\scriptsize{0.00}} & {\scriptsize{0.00}} & {\scriptsize{0.00}} & {\scriptsize{0.00}} & {\scriptsize{0.00}} & {\scriptsize{0.00}} & {\scriptsize{0.00}}\tabularnewline
 & \textit{\scriptsize{Med.}} & {\scriptsize{60.99}} & {\scriptsize{136.99}} & {\scriptsize{26.00}} & {\scriptsize{49.00}} & {\scriptsize{11.99}} & {\scriptsize{21.00}} & {\scriptsize{17.00}} & {\scriptsize{33.50}} & {\scriptsize{5.00}} & {\scriptsize{5.00}}\tabularnewline
 & \textit{\scriptsize{Mean}} & {\scriptsize{117.32}} & {\scriptsize{295.71}} & {\scriptsize{59.59}} & {\scriptsize{131.20}} & {\scriptsize{37.91}} & {\scriptsize{69.31}} & {\scriptsize{124.74}} & {\scriptsize{338.86}} & {\scriptsize{21.66}} & {\scriptsize{41.34}}\tabularnewline
 & \textit{\scriptsize{Max.}} & {\scriptsize{834.00}} & {\scriptsize{2,108.00}} & {\scriptsize{1,356.00}} & {\scriptsize{2,488.00}} & {\scriptsize{773.00}} & {\scriptsize{1,793.00}} & {\scriptsize{3,928.99}} & {\scriptsize{11,522.99}} & {\scriptsize{867.00}} & {\scriptsize{1,553.00}}\tabularnewline[1ex]
\multirow{4}{*}{\textbf{\scriptsize{MS}}} & \textit{\scriptsize{Min.}} & {\scriptsize{0.00}} & {\scriptsize{0.00}} & {\scriptsize{0.00}} & {\scriptsize{0.00}} & {\scriptsize{0.00}} & {\scriptsize{0.00}} & {\scriptsize{0.00}} & {\scriptsize{0.00}} & {\scriptsize{0.00}} & {\scriptsize{0.00}}\tabularnewline
 & \textit{\scriptsize{Med.}} & {\scriptsize{0.80}} & {\scriptsize{0.71}} & {\scriptsize{0.24}} & {\scriptsize{0.16}} & {\scriptsize{0.73}} & {\scriptsize{0.66}} & {\scriptsize{0.84}} & {\scriptsize{0.74}} & {\scriptsize{0.50}} & {\scriptsize{0.45}}\tabularnewline
 & \textit{\scriptsize{Mean}} & {\scriptsize{0.71}} & {\scriptsize{0.64}} & {\scriptsize{0.29}} & {\scriptsize{0.24}} & {\scriptsize{0.60}} & {\scriptsize{0.56}} & {\scriptsize{0.72}} & {\scriptsize{0.66}} & {\scriptsize{0.44}} & {\scriptsize{0.41}}\tabularnewline
 & \textit{\scriptsize{Max.}} & {\scriptsize{1.00}} & {\scriptsize{1.00}} & {\scriptsize{1.00}} & {\scriptsize{1.00}} & {\scriptsize{1.00}} & {\scriptsize{1.00}} & {\scriptsize{1.00}} & {\scriptsize{1.00}} & {\scriptsize{1.00}} & {\scriptsize{1.00}}\tabularnewline
 \bottomrule
\end{tabular}
\end{table*}

\subsection{Experimental Environment}

All the experiments were performed on a quad-core Intel Xeon processor (3.1GHz) with 8GB of RAM and running Ubuntu 14.04.3 LTS (Trusty Tahr).



\subsection{Employed Tools}

We use PIT, a popular mutation testing tool, to support our experiments. We use the 1.1.5 release with the extended set of mutants that it supports. To enable a comparison with the comprehensive mutants, we modified PIT to support them. The next section details all the considered mutants. 


%
%
%
%
%


\subsection{Employed Mutants}


The employed mutant sets are described in Table \ref{tab:ops}. The common mutants are described in the upper part of the table while the comprehensive ones are described in the lower part. The comprehensive mutant set was formed based on the beliefs of the mutation testing community \cite{MutKeynote2015} and the literature. In particular we adapt to Java the set of mutants that was suggested and used in the following studies \cite{OffuttLRUZ96, AndrewsBLN06, 0020331}.

Note that the comprehensive set of mutants includes all the common ones. For each mutant, Table \ref{tab:ops} records its name, a description of the transformation performed and an example.

Special care was taken in order to reduce the duplicated mutant instances \cite{PapadakisJHT15} by removing the overlap between the operators. It is also possible that some mutants might be redundant \cite{JustKS12}. However, using an analysis similar to \cite{JustKS12} may degrade the effectiveness of the method in cases of mutants that cannot be propagated. We discuss this issue in the related work section, i.e., \ref{sec:rwacc}. To avoid such risk we rely on disjoint mutants to remove redundancies among the mutants. 

Table \ref{operators_table} presents some descriptive statistics (minimum median, mean and maximum values) about the employed mutants as they appear in the classes of the studied projects. Thus, the table records details about the number of mutants, the number of killable mutants (determined based on the available test suite), and mutation score for the common and comprehensive operators for each project.

\subsection{Analysis Procedure for Answering the Research Questions}




To answer RQ1, we constructed test suites using the common set. This was performed by incrementally adding random tests in the suites and keeping only those that increase mutation score. So, if the randomly selected tests failed to kill any additional mutant, i.e., it is redundant with respect to the employed mutants, the test was not included. This is a typical process followed by many previews studies, e.g., \cite{AndrewsBLN06,OffuttPTZ96}. Thus, we measured (a) the number of mutants of the comprehensive set that are killed by the tests selected based on the common set and (b) the number of mutants of the comprehensive sets found when using all available tests. Since the tests were selected at random, this process was repeated 30 times. As a result, we obtain 30 instances for each one of the two measures for every class of each project. We compared them with 
Wilcoxon test using the R statistical computing project \cite{r}. From this test, we obtain a p-value which represents the probability that measure (a) is higher than measure (b). Following the usual statistical inference procedures we consider the differences as statistically significant if they provide a p-value lower than 0.05, i.e., 5\% is our significance level.  
In RQ1, we record the number of classes for which there is a statistically significance difference. 
The ratio (a)/(b) forms the objective comparison score when using all mutants. The values of Table \ref{operators_table} imply that the projects have classes with only 1 mutant and 0 killable ones. Also, there are classes where all mutants are killed. Thus, we base the results of the objective comparison at the class granularity level and present them according the first three quartiles, i.e., according to the ordered 25\%, median (50\%) and 75\% values. Finally, we compute the ratio of the disjoint mutants of the comprehensive set that are found by the tests selected based on the common set. The distance from value 1 on both the objective comparison and disjoint mutants scores quantify the effectiveness differences between the examined mutants. 


To answer to RQ2, we measure the easiness of killing mutants. The easiness of killing a mutant is defined as the number of test cases that kill a mutant, towards the total number of test cases. As a result, when 100\% of the test cases kill a mutant, the latter is denoted as very easy to kill.

To answer RQ3, we applied mutation analysis as it is supported by the current version of the tool and record the time required.

\section{Results \& Answers to the Research Questions}\label{xp_results}

\begin{figure*}[!t]
\centering

  \includegraphics[width=0.385\textwidth]{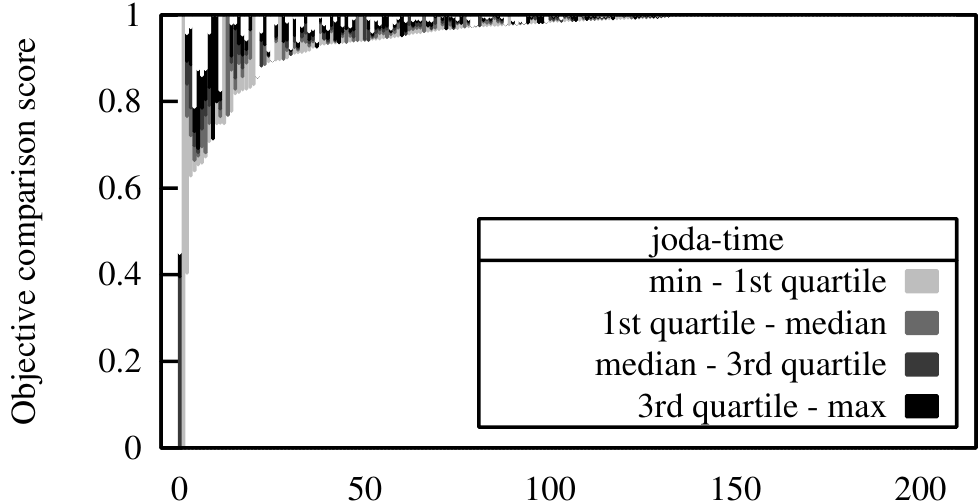}
 \hspace{3em}
  \includegraphics[width=0.385\textwidth]{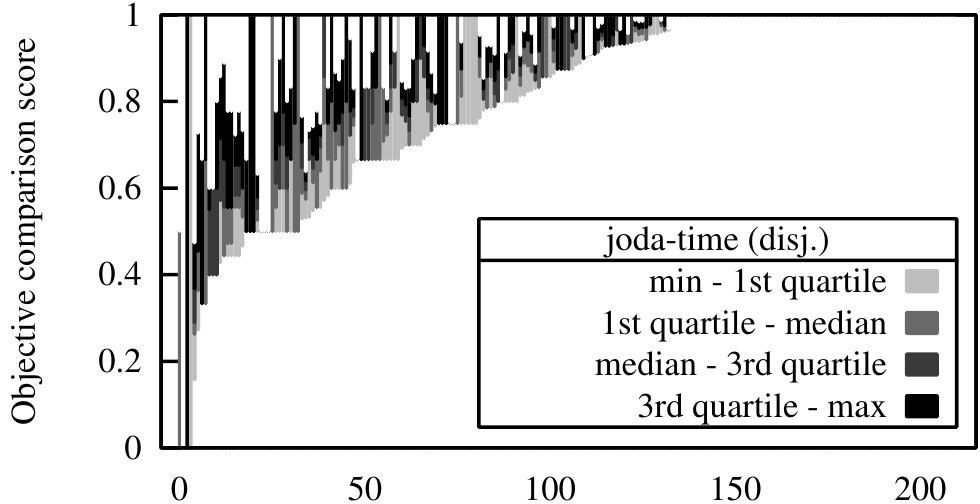}

  \vspace{.5em}
  \includegraphics[width=0.385\textwidth]{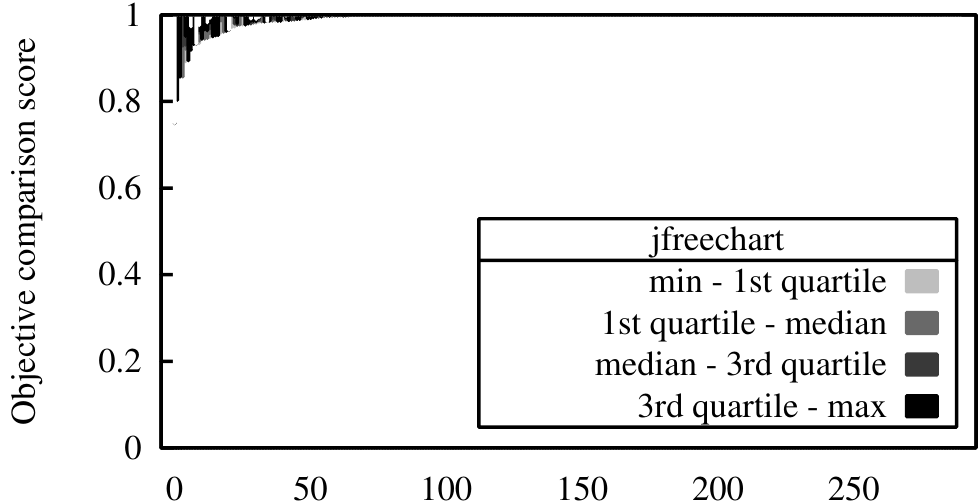}
 \hspace{3em}
  \includegraphics[width=0.385\textwidth]{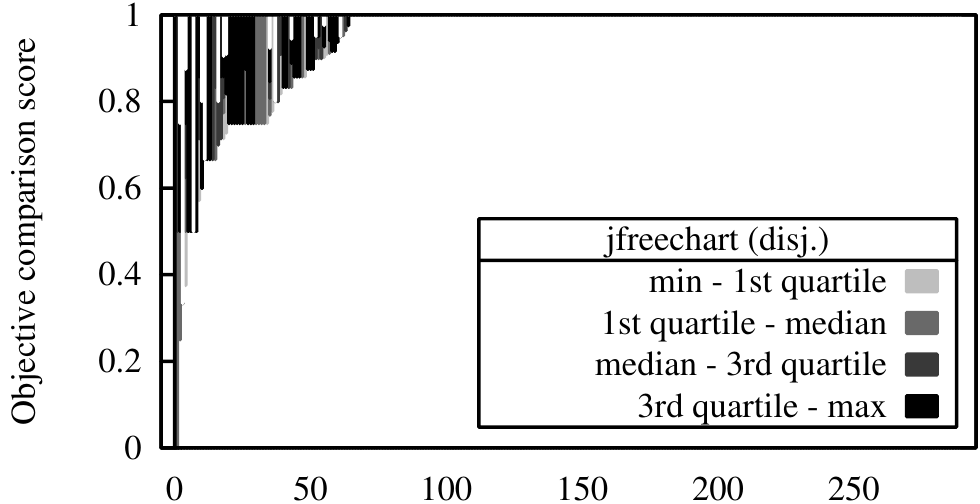}

  \vspace{.5em}

  \includegraphics[width=0.385\textwidth]{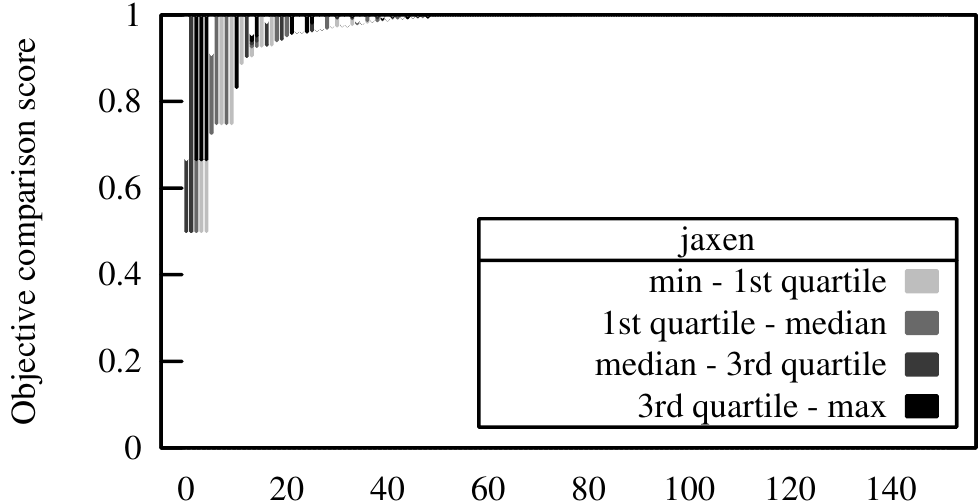}
 \hspace{3em}
  \includegraphics[width=0.385\textwidth]{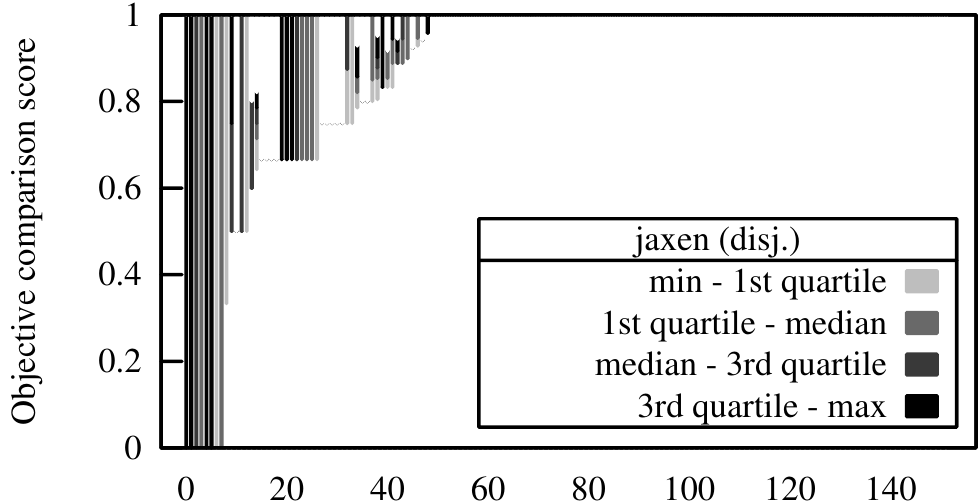}

 \vspace{.5em}

  \includegraphics[width=0.385\textwidth]{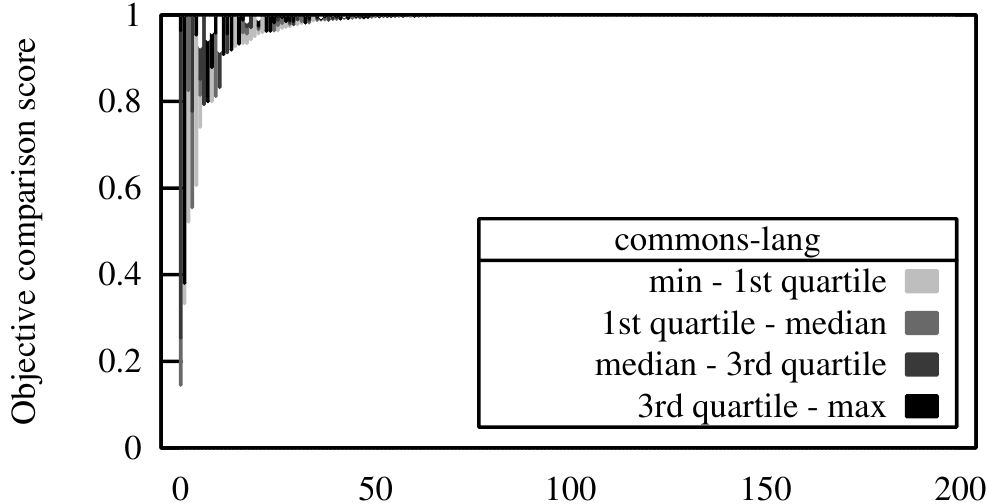}
 \hspace{3em}
  \includegraphics[width=0.385\textwidth]{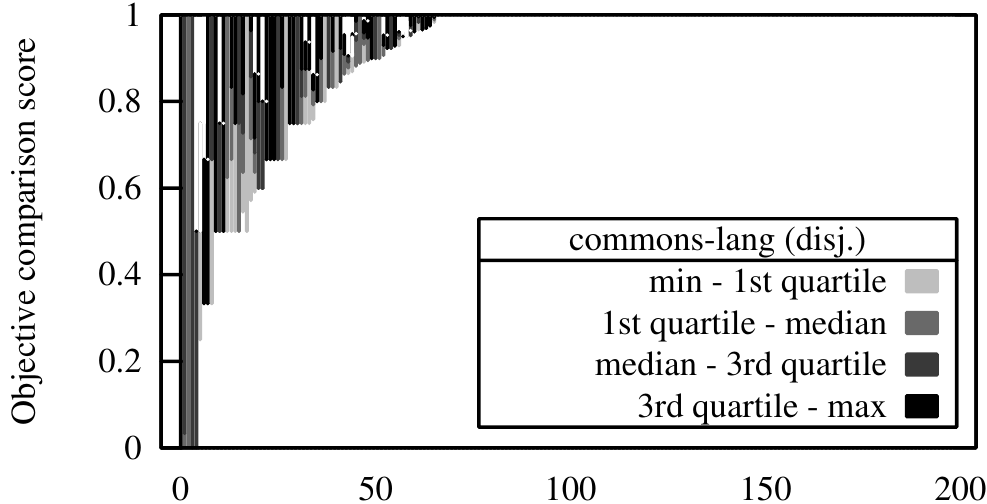}

  \vspace{.5em}

  \includegraphics[width=0.385\textwidth]{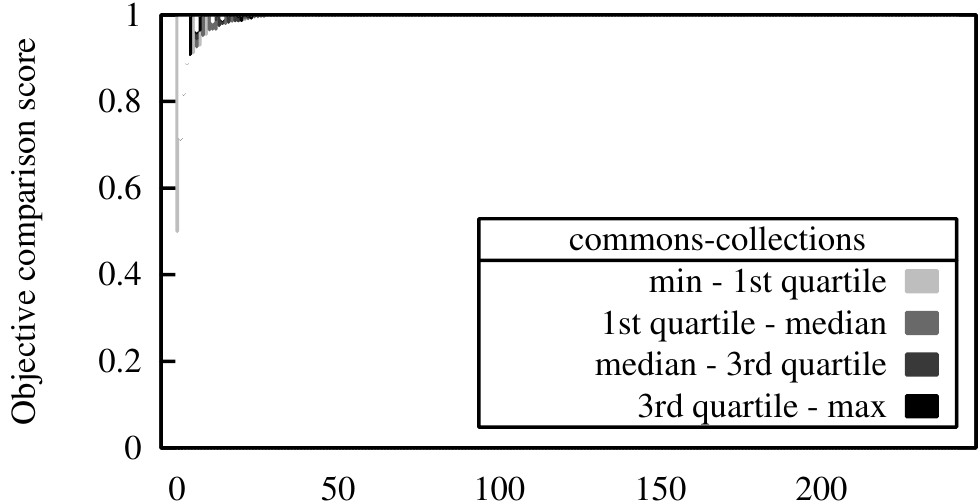}
 \hspace{3em}
  \includegraphics[width=0.385\textwidth]{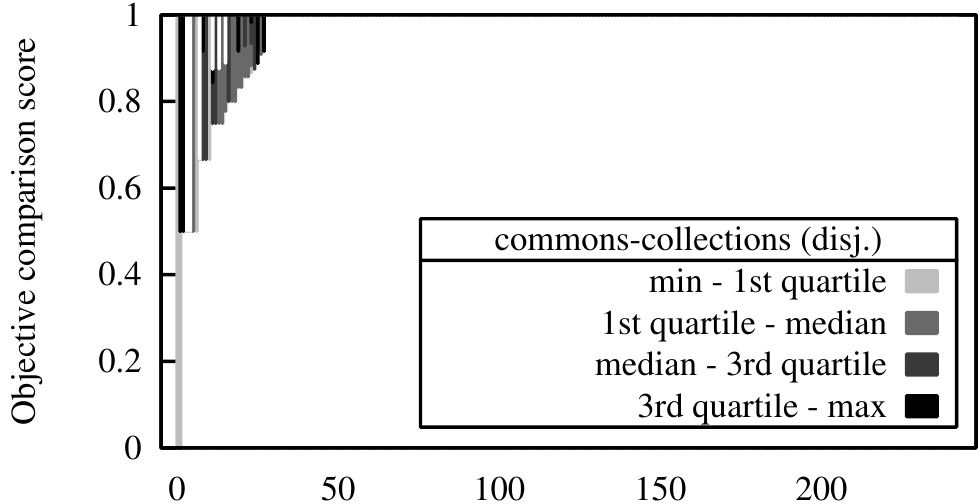}

\caption{Objective comparison results (RQ1 - effectiveness). The plots on the left side display the results of all mutants while on the right side the results of the disjoint mutants. The y-axis represents the distribution of the 30 scores per class, i.e., the minimum, first quartile, median, third quartile and maximum, while the x-axis represents the Java classes of the programs.}
\label{objectivecomp}
\end{figure*}

\begin{figure*}[!t]
\centering

  \includegraphics[width=0.385\textwidth]{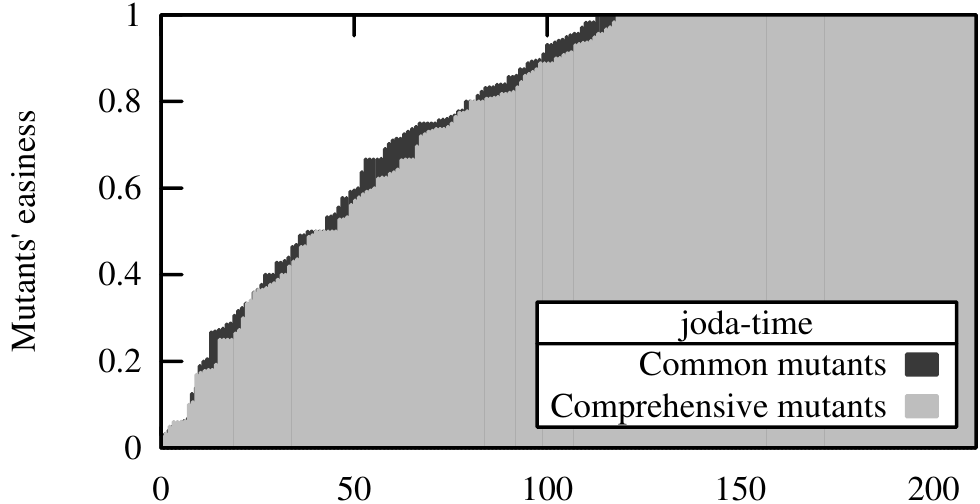}
 \hspace{3em}
  \includegraphics[width=0.385\textwidth]{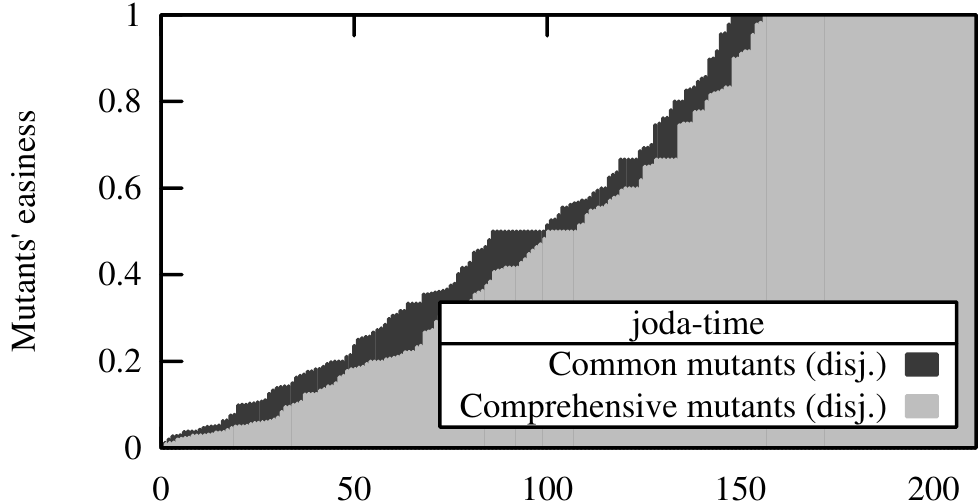}
  \vspace{.5em}

  \includegraphics[width=0.385\textwidth]{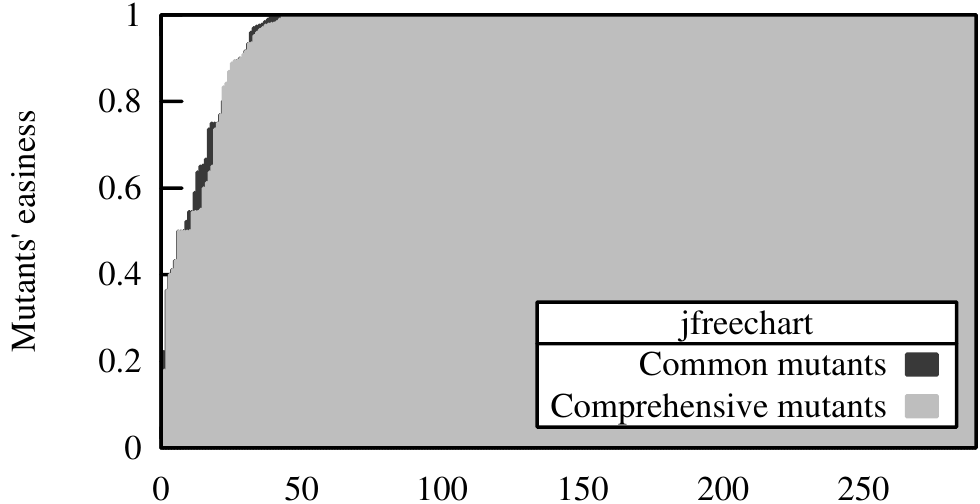}
 \hspace{3em}
  \includegraphics[width=0.385\textwidth]{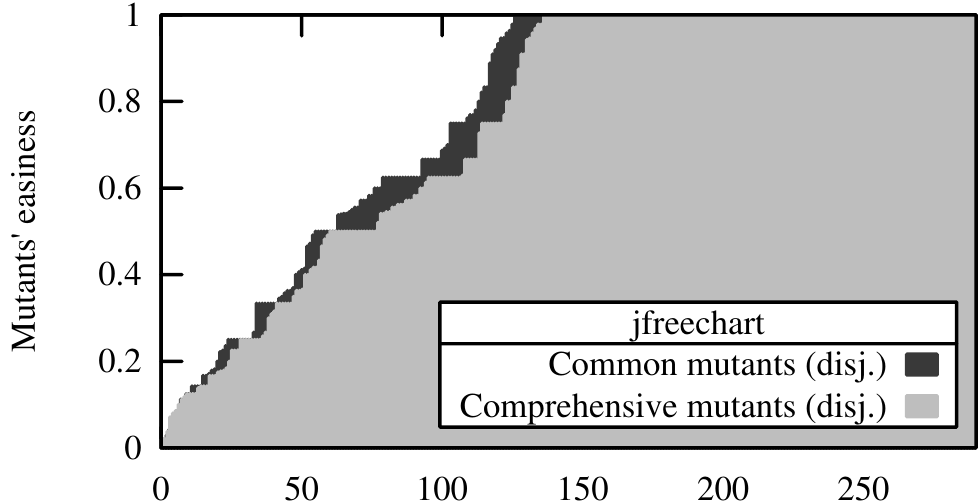}

  \vspace{.5em}

  \includegraphics[width=0.385\textwidth]{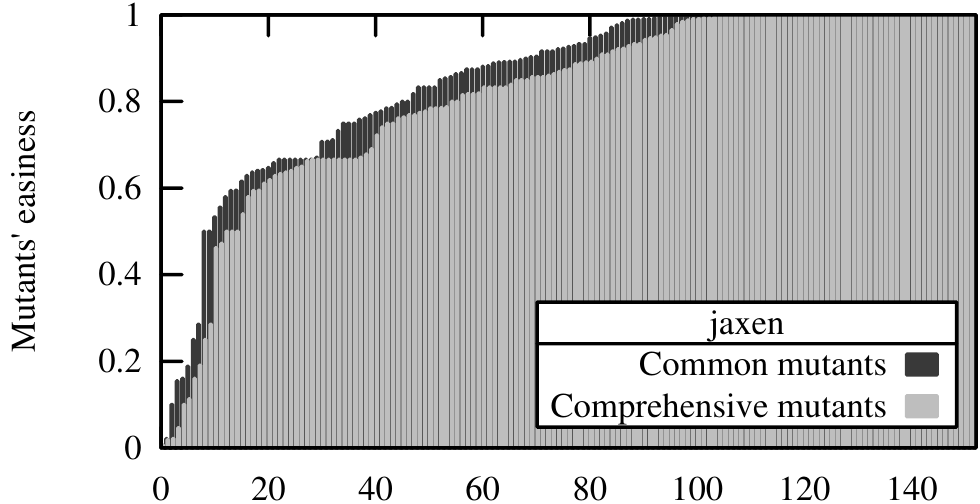}
 \hspace{3em}
  \includegraphics[width=0.385\textwidth]{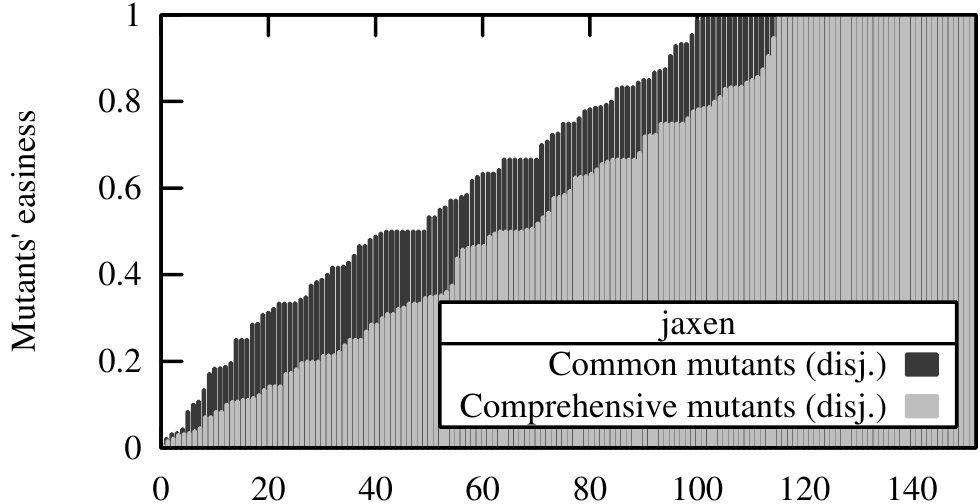}
 
 \vspace{.5em}

  \includegraphics[width=0.385\textwidth]{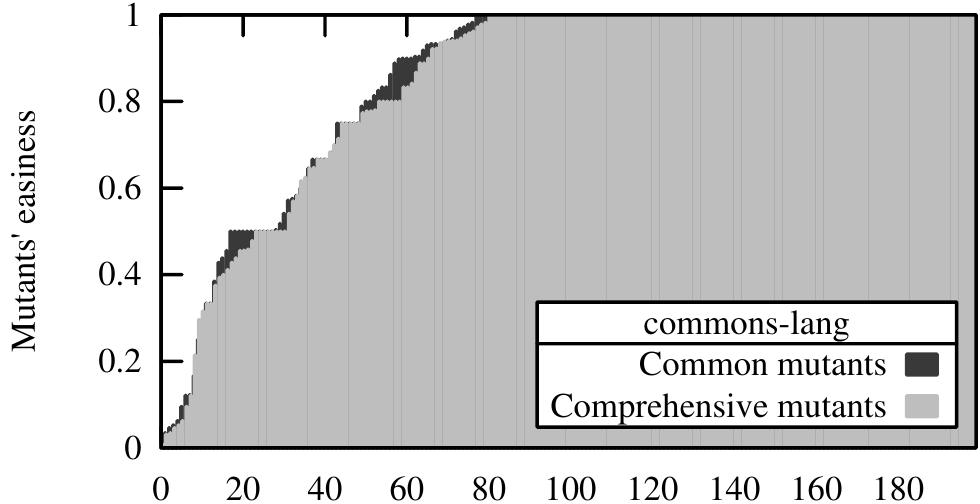}
 \hspace{3em}
  \includegraphics[width=0.385\textwidth]{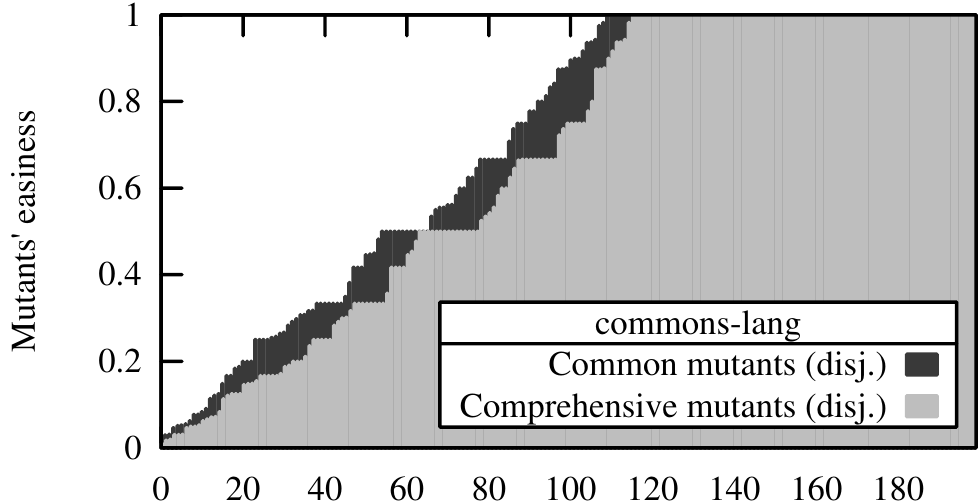}
  \vspace{.5em}

  \includegraphics[width=0.385\textwidth]{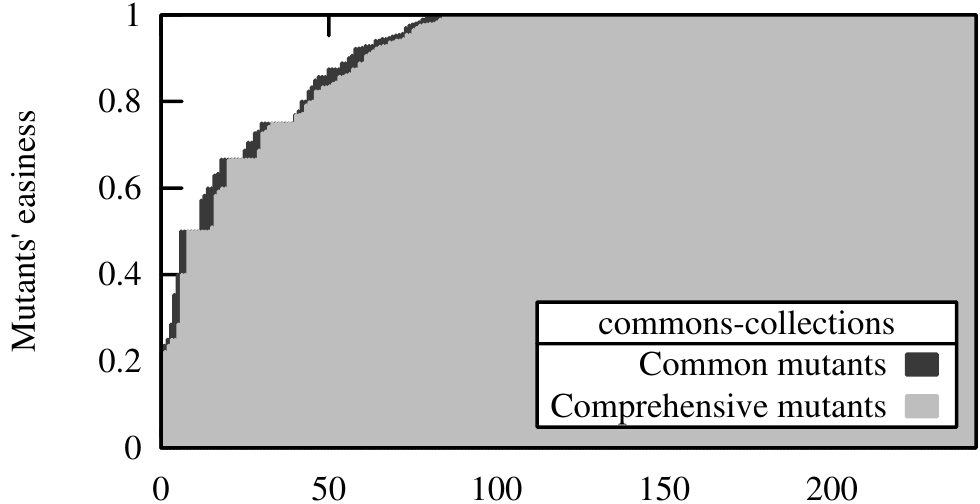}
 \hspace{3em}
  \includegraphics[width=0.385\textwidth]{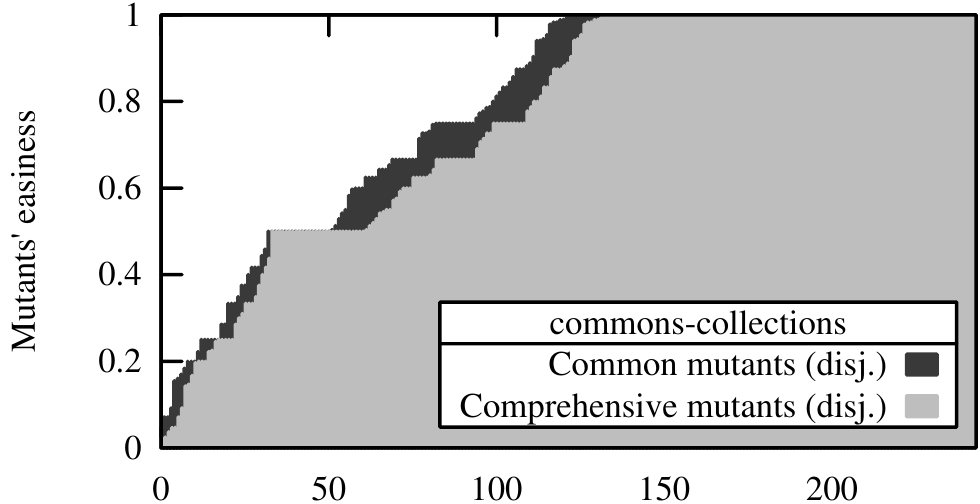}

\caption{Easiness of killing mutants (RQ2 - easiness). The plots on the left side display the easiness of all mutants while on the right side they display the easiness of the disjoint mutants. The y-axis represents the median mutant easiness, while x-axis represents the Java classes of the programs.}
\label{easiness}
\end{figure*}

This section reports on the experimental results and answers the RQs stated in previous section.

\subsection{RQ1 - Effectiveness}\label{answ:rq1}

For this question, we first consider the number of comprehensive mutants killed by test cases selected based on the common mutants, and the number of comprehensive mutants killed by test cases targeting them. This forms two mutation scores for each Java class of the considered projects. The comparison has been performed 30 times, thus yielding 30 mutation scores of the two types per class. 

Table \ref{wilcox_table} records the classes for which there is a statistical significance between the two measures. As can be seen, there is a significant difference for 11\% of the classes for commons-collections, for 62\% of them for joda-time.

\begin{table}[!t]
 \caption{Number and proportion of classes for which there is statistical significance between the common and comprehensive mutant sets.}
\label{wilcox_table}
\centering


\setlength{\tabcolsep}{14.5px}


\begin{tabular}{lr}
\toprule
\textbf{\scriptsize{Subject}} & \textbf{\scriptsize{\#Classes (proportion)}}\tabularnewline
\midrule
{\scriptsize{joda-time}} & {\scriptsize{130 (62\%)}}\tabularnewline
{\scriptsize{jfreechart}} & {\scriptsize{64 (22\%)}}\tabularnewline
{\scriptsize{jaxen}} & {\scriptsize{43 (28\%)}}\tabularnewline
{\scriptsize{commons-lang}} & {\scriptsize{63 (32\%)}}\tabularnewline
{\scriptsize{commons-collections}} & {\scriptsize{26 (11\%)}}\tabularnewline
\bottomrule
\end{tabular}

\end{table}

We now evaluate the percentage of mutants of the comprehensive set that are killed by the test cases selected based on the common set. This percentage is denoted as objective comparison score. Figure \ref{objectivecomp} records the objective comparisons scores for the 5 programs. The objective comparison has been performed 30 times per class (represented on the x-axis), thus yielding 30 different scores for each class of each program.  

The 4 colors of the plot represent the distribution of the 30 ordered scores according to the quartiles. Thus, from the lightest to the darkest color, the first, second, third and fourth 25\% of the resulting scores are represented. For instance, a light gray bar (the first quartile reaching 0.6) means that the lowest 25\% of the 30 scores obtained are below or equal to 0.6. The black area represents the values above the third quartile, i.e., last 25\% of the 30 scores. In that, if a bar is completely light gray, it means that most of the mutants killed by the test cases are the same on both sets, while the presence of darker colors on the graph indicates that there are mutants that are missed by the test cases.

The plots on the left  part of Figure \ref{objectivecomp} represent the scores when considering all the mutants while the plots on the right side depict the results when considering the disjoint mutants. On the left side of the figure, i.e., when all the mutants are considered, we can see that in most of the cases the scores are close to 1, which means that there are only a few mutants missed by the test cases. This is especially the case for jfreechart and commons-collections. For the other projects, there is a larger proportion of classes with lower scores, indicating that there are more mutants of the comprehensive sets missed by the test cases. Considering now the right part of Figure \ref{objectivecomp} that concerns the disjoint mutants, we can see that the proportion of missed mutants is even more important, in particular for joda-time where mutants are missed in more than half of the classes.

To conclude, there is a significant difference between the common mutants and the comprehensive ones. This shows that the test cases miss many killable mutants from the comprehensive set. The difference is even more significant  when considering disjoint mutants.


\subsection{RQ2 - Easiness}

Here, we evaluate whether mutants are difficult to kill or not. The easiness of killing a mutant is the percentage of the test cases that kill this mutant. Thus, if all the test cases kill a given mutant, the easiness is 1. By contrast, an easiness close to 0 means that the mutant is very difficult to kill, since only few test cases are able to identify it.

Figure \ref{easiness} shows the easiness, median values, of the common mutants against the comprehensive ones for each class. A greater surface indicates that the corresponding mutants are easier to kill. The plots on the left part consider all the mutants while the right side are for the disjoint mutants. From these results, we can observe that the comprehensive mutants (represented by the gray bars) are harder to kill. For all the mutants, the difference in terms of easiness compared to the common mutants range from 2-5\% for the 50\% of the commons-lang program classes. In the case of jaxen, the easiness difference is 12\% in almost the 60\% of the program classes. With respect to the disjoint mutants, the difference goes beyond 20\% for approximately 75\% of the jaxen classes.

To summarize, the comprehensive mutants are harder to kill than the common ones when considering either all the mutants or the disjoint ones only. It means that the comprehensive set introduces faults which are more difficult to expose.

\begin{table*}[!t]

\caption{Execution time in seconds for the common and comprehensive mutants (RQ3 - scalability).}
\label{exec_time}
\centering


\setlength{\tabcolsep}{9px} 

{\scriptsize{}}%
\begin{tabular}{lrrrrrrrrr}
\toprule
 & \multicolumn{3}{c}{\textbf{\scriptsize{Common set}}} &  &  & \multicolumn{4}{c}{\textbf{\scriptsize{Comprehensive set}}}\tabularnewline
\textbf{\scriptsize{Subjects}} & \textbf{\scriptsize{Mutants}} & \textbf{\scriptsize{Killable}} & \textbf{\scriptsize{Time}} & \textbf{\scriptsize{Time/Mutant }} &  & \textbf{\scriptsize{Mutants}} & \textbf{\scriptsize{Killable}} & \textbf{\scriptsize{Time / Overhead}} & \textbf{\scriptsize{Time/Mutant}}\tabularnewline
\cmidrule(l{5pt}r{40pt}){1-1}\cmidrule(l{5pt}r{5pt}){2-5}\cmidrule(l{5pt}r{5pt}){7-10}
{\scriptsize{joda-time 2.8.1}} & {\scriptsize{35,297}} & {\scriptsize{25,224}} & {\scriptsize{1,138}} & {\scriptsize{0.032}} &  & {\scriptsize{99,343}} & {\scriptsize{63,578}} & {\scriptsize{3,531 / 210\%}} & {\scriptsize{0.035}}\tabularnewline
{\scriptsize{jfreechart 1.0.19}} & {\scriptsize{81,960}} & {\scriptsize{22,289}} & {\scriptsize{2,398}} & {\scriptsize{0.029}} &  & {\scriptsize{256,370}} & {\scriptsize{49,069}} & {\scriptsize{6,589 / 175\%}} & {\scriptsize{0.026}}\tabularnewline
{\scriptsize{jaxen 1.1.6}} & {\scriptsize{14,334}} & {\scriptsize{7,014}} & {\scriptsize{1,221}} & {\scriptsize{0.085}} &  & {\scriptsize{34,960}} & {\scriptsize{12,823}} & {\scriptsize{31,077 / 2,445\%}} & {\scriptsize{0.889}}\tabularnewline
{\scriptsize{commons-lang 3 3.4}} & {\scriptsize{34,502}} & {\scriptsize{27,443}} & {\scriptsize{2,803}} & {\scriptsize{0.081}} &  & {\scriptsize{100,553}} & {\scriptsize{74,550}} & {\scriptsize{8,023 / 186\%}} & {\scriptsize{0.080}}\tabularnewline
{\scriptsize{commons-collections 4 4.0}} & {\scriptsize{24,308}} & {\scriptsize{8,449}} & {\scriptsize{570}} & {\scriptsize{0.023}} &  & {\scriptsize{49,354}} & {\scriptsize{16,126}} & {\scriptsize{1,230 / 116\%}} & {\scriptsize{0.002}}\tabularnewline

\bottomrule
\end{tabular}

\end{table*}

\subsection{RQ3 - Scalability}

To evaluate the cost of using the comprehensive mutants, we measure the execution time of both the common and comprehensive sets. Table \ref{exec_time} records the number of mutants, killable mutants, execution time in seconds and execution time per mutant in seconds for both sets of mutants.

Considering the time in seconds, the comprehensive mutants require more time than the common ones. This is natural since the number of mutants is also much higher. The highest execution time is for jaxen with 31,077 seconds, i.e., approximately 8 hours and a half. Focusing on the time per mutants, the execution times between the two sets become rather similar, except for jaxen where the execution time is approximately 10 times longer.

Overall, the comprehensive set of mutants has an overhead in terms of execution time, but this is not unbearable.


%
%
\section{Related Work}\label{related}

The following sections present studies with respect to mutation testing (\ref{sec:rwma}), mutant reduction techniques (\ref{sec:rwmr}) and the accuracy of mutation score (\ref{sec:rwacc}). 

\subsection{Mutation Analysis}\label{sec:rwma}

Mutation analysis was initially introduced as a test method helping the generation of test cases \cite{DeMillo:1978}.  However, in recent years it has been proven to be quite powerful and capable of supporting various software engineering tasks \cite{Offutt11}. In particular mutants have been used to guide test generation \cite{PapadakisBM10, FraserZ12}, test oracle generation \cite{FraserZ12}, to assist the debugging activities \cite{PapadakisT15, NicaNW13}, to evaluate fault detection ability \cite{AndrewsBLN06} and to support regression testing activities like test selection and prioritization \cite{ShiGGZM14, ZhangMZK12}. The method has also been applied to models \cite{AichernigAJKKSS14}, software product lines \cite{HenardPPKT13} and combination strategies \cite{PapadakisHT14}.

One of the main issues of the method is equivalent mutants \cite{HieronsHD99, MadeyskiOTJ14}. Despite the efforts from the community, e.g., \cite{5487526}, and recent advances \cite{KintisPM15, PapadakisJHT15, BardinDDKPTM15}, the problem remains open \cite{PapadakisJHT15}. Similar situation arises when considering mutation-based test generation \cite{PapadakisBM10, FraserZ12}.

Mutation has become popular \cite{5487526} thanks to its ability to represent real faults \cite{AndrewsBLN06}. Also, many modern mutation tools are integrated with build systems and development tools, thus making their application easy \cite{DelahayeB15}. However, such tools usually support mutants that are more restrictive than the popular set, and hence, overestimate or underestimate the employed measures, as shown by the present paper. Previous research suggested that the comprehensive mutant set has the ability to prod-subsume and to reveal more faults than most of the other white-box testing criteria \cite{Offutt11, OffuttPTZ96}.

\subsection{Mutant Reduction}\label{sec:rwmr}

From the early days of mutation testing, it was evident that mutants were far too numerous to be used in practice. Therefore, researchers have tried to identify subsets of them that are representative. The first reduction was made towards the coupling effect hypothesis, which states that tests revealing simple mutants can also reveal complex ones 
\cite{DeMillo:1978, Offutt89}. 

A straight way to reduce the number of mutants is to randomly sample them \cite{PapadakisM10}. Although, sampling can provide a range of trade-offs, Papadakis and Malevris \cite{PapadakisM10} provided evidence that mutant sampling ratios of 10\% to 60\% have a loss on fault detection from 26\% to 6\%, respectively. Similar results have been shown in the study of Wong \etal \cite{WongM95}.

Other mutant reduction strategies fall in the category of selective mutation \cite{OffuttLRUZ96}. Selective mutation tries to reduce the arbitrariness of random sampling by using only specific types of mutants. 

\subsection{Duplicated, Trivial and Redundant Mutants}\label{sec:rwacc}


The presence of redundant mutants has long been recognized, i.e., since 1993 \cite{Tai93}, but only recently, received attention. The studies of Tai \cite{Tai06, Tai93} were focused on reducing the application cost of fault-based testing strategies. This was based on constraints that restrict the mutants introduced by the relational and logical operators so that they only consider non-redundant ones. Thus, their suggestion was to restrict the mutant instances of the relational and logical operators to the minimum possible number. In a later study, Kaminski \etal \cite{KaminskiAO13, KaminskiAO11} came to a similar conclusion about the relational operator. Thus, they suggested that mutation testing tools should reduce the number of redundant mutants by restricting the mutant instances of the relational operator.

Papadakis and Malevris \cite{PapadakisM12} suggested using minimized constraint mutant instances  to efficiently generate mutation-based test cases. Thus, when aiming at generating test cases there is no point in aiming at non-redundant mutants. On the same lines, Just \etal  \cite{JustKS12}, demonstrated that by constraining the relational and logical operators, it is possible to reduce some of the redundancy between the mutants. 

All the approaches discussed so far were based on an analysis at the predicate level, specifically designed for ``weak'' mutation. Thus, when applied to strong mutation, the analysis may not hold. This can be due to the following two reasons; a) constructs like loops and recursion can make a statement to be exercised multiple times, and b) these approaches assume that when the identified mutants are killed the redundant ones are also killed. However, it is likely that the identified mutants are equivalent while the non-redundant ones are not, thus, resulting in degradation in the effectiveness of the method. These issues motivated the use of disjoint mutants that do not suffer from these problems.

The first study suggesting the use of non-redundant mutants when comparing testing techniques is that of Kintis \etal \cite{KintisPM10}. As discussed in Section \ref{sec:disjoint}, Kintis \etal introduced the notion of disjoint mutants and demonstrated that the majority of mutants produced by the MuJava mutation testing tool is redundant. 
In the same lines, Amman \etal \cite{AmmannDO14} defined algorithms for generating a minimum set of mutants based on the notion of dynamic subsumption. Their results confirmed the findings of Kintis \etal by demonstrating that the majority of mutants used by the MuJava and Proteum mutation testing tools are redundant. Later, Kurtz \etal \cite{KurtzAO15} used static analysis techniques, such as symbolic execution to identify the minimum set of mutants. 

Recently Papadakis \etal \cite{PapadakisJHT15} used compilers to eliminate duplicated mutants, a special form of mutant redundancy. Duplicated mutants are those that are mutually equivalent but differ from the original program. In the study of Papadakis \etal \cite{PapadakisJHT15} it is reported that 21\% of all mutants is duplicated and can be removed by using compiler optimization techniques.

All the approaches discussed in this section either identified the problem of trivial/redundant mutants or used some form of redundancy elimination. However, none of them studied the differences of the commonly used operators from those suggested by the literature. Additionally, none of them uses disjoint mutants on real-world programs.

\section{Conclusions}\label{conclusion}
This paper investigates the extent to which mutants used by popular mutation testing tools like PIT conform to mutation testing standards. Our study revealed a large divergence in the effectiveness of the popular mutants from the comprehensive ones. Comprehensive mutants are not only harder to kill but also score considerably higher than the common ones  in most of the examined cases. Additionally, we report results by considering both concepts of disjoint and mutant easiness. Thus, we point out the importance of the problems introduced by both trivial and redundant mutants to be considered in future evaluations.


%
%
%



%

\balance
\bibliographystyle{IEEEtran}
\bibliography{IEEEabrv,biblio}

\end{document}